\documentclass[twocolumn,showpacs,preprintnumbers]{revtex4}
\usepackage{amssymb}
\usepackage{amsmath}
\usepackage{graphicx}
\usepackage{dcolumn}
\usepackage{bm}
\usepackage{color}

\begin{document}

\title{Rich phase diagram of quantum phases in the anisotropic subohmic spin-boson model}
\author{Yan-Zhi Wang$^{1}$, Shu He$^{2}$, Liwei Duan$^{1}$, and Qing-Hu Chen$%
^{1,3,*}$}

\address{
$^{1}$ Zhejiang Province Key Laboratory of Quantum Technology and Device, Department of Physics, Zhejiang University, Hangzhou 310027, China \\
$^{2}$ Department of Physics and Electronic Engineering,
Sichuan Normal University, Chengdu 610066, China\\
$^{3}$  Collaborative Innovation Center of Advanced Microstructures, Nanjing University, Nanjing 210093, China
 }\date{\today }

\begin{abstract}
We study the anisotropic spin-boson model (SBM) with the subohmic bath by a numerically exact
method based on variational matrix product states. A rich phase diagram is
found in the anisotropy-coupling strength plane by calculating several
observables. There are three distinct quantum phases: a delocalized phase with
even parity (phase I), a delocalized phase with odd parity (phase II), and
a localized phase with broken $Z_2$ symmetry (phase III), which intersect at a
quantum tricritical point. The competition between those phases would give
overall picture of the phase diagram. For small power of the spectral
function of the bosonic bath, the quantum phase transition (QPT) from phase
I to III with mean-field critical behavior is present, similar to the
isotropic SBM. The novel phase diagram full with three different phases can
be found at large power of the spectral function: For highly anisotropic
case, the system experiences the QPTs from phase I to II via 1st-order, and
then to the phase III via 2nd-order with the increase of the coupling
strength. For low anisotropic case, the system only experiences the
continuous QPT from phase I to phase III with the non-mean-field critical
exponents. Very interestingly, at the moderate anisotropy, the system would
display the continuous QPTs for several times but with the same critical
exponents. This unusual reentrance to the same localized phase is discovered
in the light-matter interacting systems. The present study on the anisotropic SBM could open an
 avenue to  the rich quantum criticality.
\end{abstract}

\maketitle

\section{Introduction}

The quantum phase transition (QPT) has been studied for many years and continues
to be a hot topic in many correlated matters and  light-matter
interacting systems \cite{Sachdev}, such as  fermionic \cite{hubbard},
spin \cite{Sachdev}, bosonic \cite{bose}, as well as  fermion
(spin)-boson coupling systems \cite{HubHol,Leggett}. Because fermions have
both  spin and charge degree of freedoms,  rich and novel quantum
phases can emerge in the fermionic model and the bosonic model if bosons are
formed by composite fermions or cold atoms in  strongly correlated systems.

In  light-matter interacting systems, many prototype models including the
quantum Rabi model \cite{QRM}, the Dicke model \cite{Dicke}, and the
spin-boson model (SBM)\cite{Leggett} only experience a single QPT from the
normal to supperradiant phase for the single mode bosonic cavity or
delocalized to localized phase for the bosonic bath. The QPTs of most models
are trivially of the mean-field nature. Only the sub-Ohmic SBM can also
display the non-mean-field critical behavior with the large power of the
spectral function of the bosonic bath \cite{Bulla}. The nonclassical
critical behavior is at the heart of so-called local quantum criticality
\cite{si}.

To obtain the rich phase diagram, the generalized Dicke models, such as the
anisotropic Dicke model ~\cite{yejw,ciute}, the anisotropic Dicke model with
 Stark coupling terms~ \cite{zhiqiang}, and the isotropic Dicke model
with antiferromagnetic bias fields ~\cite{puhan} have been recently studied
by several groups. A quantum tricritical point (QuTP)~\cite{Griffiths} is
seldomly supported in  solid-state materials, and is almost impossible to
appear in  prototype models of  light-matter interacting systems due
to the single phase transition. Interestingly, it has been found to exist in the
anisotropic Dicke model ~\cite{ciute} and the isotropic Dicke model with a
special configuration of bias fields ~\cite{puhan}. In the former model, the
QuTP lies at the symmetric line of the superradiant \textquotedblleft
electric\textquotedblright\ and \textquotedblleft
magnetic\textquotedblright\ phases which can be switched mutually by
interchanging the interaction terms with the two quadratures of a bosonic
mode. While in the latter model, the 1st-order critical line meets the
2nd-order one at the QuTP. Yet it has not been found that three critical
lines intersect at the QuTP and separate three phases in an asymmetric way
as in the $He^{3}-He^{4}$ mixture~\cite{Griffiths} in the light-matter
interacting systems until now, to the best of our knowledge.

The phase diagrams in those generalized Dicke models become richer than
their prototype models, but still only include one 1st-order and one
2nd-order critical lines, possibly due to the fact that only a single phase
transition with mean-field type is present in the prototype models. This
situation might be changed in a generalized model if its prototype one can
exhibit both non-mean-field and mean-field critical behaviors, like the
subohmic SBM.

As is well known that the SBM is a paradigmatic model in many fields,
ranging from quantum optics \cite{Scully}, to condensed matter physics \cite%
{Leggett}, to open quantum systems \cite{Breuer,weiss1}. \ With the advance
of modern technology, various qubit and oscillator coupling systems can be
engineered in many solid-state devices, such as superconducting circuits
\cite{Niemczyk,Yoshihara}, cold atoms \cite{Dimer}, and trapped ions \cite%
{Cirac}. Recently, the SBM has been realized by the ultrastrong coupling of
a superconducting flux qubit to an open one-dimensional transmission line ~%
\cite{Forn2}. The counterrotating terms can be suppressed in some proposed
schemes \cite{yejw,Keeling,Fanheng}. In some systems, the anisotropy appears
quite naturally, because they are controlled by different input parameters~
\cite{Grimsmo1}.

In the subohmic SBM, the 2nd-order QPT from the delocalized phase, where
spin has the equal probability in the two states, to localized phase, in
which spin prefers to stay in one of the two states, has been studied
extensively \cite%
{Bulla,QMC,ED,Vojta1,Zhang,Chin,guo2012critical,Frenzel,CRduan,HeShu1}.
Unlike the Dicke model and the quantum Rabi model, the SBM has various
universality classes, depending on the power of the spectral function of the
bosonic bath. Therefore its generalized model including anisotropy might
support richer quantum phases with the help of the additional parameter
dimension.

In this paper, we will extend the variational matrix product state (VMPS)
approach ~\cite{guo2012critical} to study the anisotropic spin-boson model
(ASBM) with the subohmic bath. The multi-coherent state (MCS) variational
approach is also employed to provide an independent evidence of  emerging
new phase. The paper is organized as follows. In Sec. II, we introduce the
ASBM briefly. Some methodologies including the VMPS and  the MCS variational
approaches are  reviewed briefly. The rich phase diagrams revealed by
the VMPS method are presented in Sec. III. A QuTP is observed and the
quantum criticality based on VMPS studies on the parity, the order
parameter, and the entanglement entropy are also analyzed. Finally,
conclusions are drawn in Sec. IV.

\section{Generalized Model Hamiltonian and methodologies}

The ASBM Hamiltonian can be written as ($\hbar =1$)
\begin{eqnarray}
\hat{H} &=&\frac{\Delta }{2}\sigma _{z}+\frac{\epsilon }{2}\sigma
_{x}+\sum_{k}\omega _{k}a_{k}^{\dag }a_{k}+\frac{1}{2}\sum_{k}g_{k}\left(
a_{k}^{\dag }+a_{k}\right) \sigma _{x}  \notag \\
&&+\frac{\lambda }{2}\sum_{k}g_{k}\left( a_{k}-a_{k}^{\dag }\right) i\sigma
_{y},  \label{Hamiltonian2}
\end{eqnarray}%
where $\sigma _{i}$ ($i=x,y,z$) are the Pauli matrices, $\Delta $ is the
qubit frequency, $\epsilon $ is the energy bias applied in a two-level
system, and $\lambda $ reflects the degree of anisotropy of this model. $%
a_{k}$ ($a_{k}^{\dag }$) is the bosonic annihilation (creation) operator
which can annihilate (create) a boson with frequency $\omega _{k}$, and $%
g_{k}$ denotes the coupling strength between the qubit and the bosonic bath,
which is usually characterized by the power-law spectral density $J(\omega )$%
,
\begin{equation}
J(\omega )=\pi \sum_{k}g_{k}^{2}\delta (\omega -\omega _{k})=2\pi \alpha
\omega _{c}^{1-s}\omega ^{s}\Theta (\omega _{c}-\omega ),  \label{spectra}
\end{equation}%
where $\alpha $ is a dimensionless coupling constant, $\omega _{c}$ is the
cutoff frequency, and $\Theta (\omega _{c}-\omega )$ is the Heaviside step
function. The power of the spectral function $s$ classifies the reservoir
into super-Ohmic $\left( s>1\right) $, Ohmic $\left( s=1\right) $, and
subohmic $\left( s<1\right) $ types. On the one hand, the isotropic SBM can
be described by Hamiltonian (\ref{Hamiltonian2}) with $\lambda =0$. On the
other hand, if the counterrotating terms involving higher excited states, $%
a_{k}^{\dag }\sigma _{+}$ and $a_{k}\sigma _{-}$, are neglected ($\lambda =1$%
), the ASBM is reduced to the SBM in the rotating-wave approximation (RWA),
which has been studied by the present authors recently~\cite{wangyz}.

The ASBM at $\epsilon=0$ possesses a $Z_{2}$ symmetry, similar to the isotropic SBM model.
The parity operator is defined as%
\begin{equation}
\hat{\Pi}=\exp \left( i\pi \hat{N}\right),  \label{parity}
\end{equation}%
where $\hat{N}=\sum_{k}a_{k}^{\dag }a_{k}+\sigma _{+}\sigma _{-}$ with $%
\sigma _{\pm }=\left( \sigma _{x}\pm i\sigma _{y}\right) /2$ is the operator
of the total excitation number. The parity operator $\hat{\Pi}$ has two
eigenvalues \ $\pm 1$, corresponding to even and odd parity in the symmetry
conserved phases. The average value of the parity may also become zero due
to the quantum fluctuations in the symmetry broken phase. So the parity can
be employed to distinguish different phases in the ASBM.

\textsl{VMPS approach.-}. To apply VMPS in the ASBM, first the logarithmic
discretization of the spectral density of the continuum bath~\cite{Bulla}
with discretization parameter $\Lambda >1$ is performed, followed by using
orthogonal polynomials as described in Ref.~\cite{Plenio_Chin}, the ASBM can
be mapped into the representation of an one-dimensional semi-infinite chain
with nearest-neighbor interaction ~\cite{Friend}. Thus, Hamiltonian (\ref%
{Hamiltonian2}) can be written as:
\begin{eqnarray}
H_{\text{chain}} &=&\frac{\Delta }{2}\sigma _{z}+\frac{\epsilon }{2}\sigma
_{x}+\frac{c_{0}}{2}(b_{0}+b_{0}^{\dag })\sigma _{x}+\lambda \frac{c_{0}}{2}%
(b_{0}-b_{0}^{\dag })i\sigma _{y}  \notag \\
&&+\sum_{n=0}^{L-2}[\epsilon _{n}b_{n}^{\dag }b_{n}+t_{n}(b_{n}^{\dag
}b_{n+1}+b_{n+1}^{\dag }b_{n})],  \label{Hamitrans}
\end{eqnarray}%
where $b_{n}^{\dag }$($b_{n}$) is the creation (annihilation) operator for a
new set of boson modes in a transformed representation with $\epsilon _{n}$
describing frequency on chain site $n$, $t_{n}$ the nearest-neighbor hopping
parameter, and $c_{0}$ the effective coupling strength between the spin and
the new effective bath. These  parameters  are expressed below{\
\begin{eqnarray*}
c_{0} &=&\sqrt{\int_{0}^{\omega _{c}}\frac{J\left( \omega \right)}{\pi}
d\omega} , \\
\epsilon _{n} &=&\xi _{s}\left( A_{n}+C_{n}\right) , \\
t_{n} &=&-\xi _{s}\left( \frac{N_{n+1}}{N_{n}}\right) A_{n},
\end{eqnarray*}%
where
\begin{eqnarray*}
\xi _{s} &=&\frac{s+1}{s+2}\frac{1-\Lambda ^{-\left( s+2\right) }}{1-\Lambda
^{-\left( s+1\right) }}\omega _{c}, \\
A_{n} &=&\Lambda ^{-j}\frac{\left( 1-\Lambda ^{-\left( j+1+s\right) }\right)
^{2}}{\left( 1-\Lambda ^{-\left( 2j+1+s\right) }\right) \left( 1-\Lambda
^{-\left( 2j+2+s\right) }\right) }, \\
C_{n} &=&\Lambda ^{-\left( j+s\right) }\frac{\left( 1-\Lambda ^{-j}\right)
^{2}}{\left( 1-\Lambda ^{-\left( 2j+s\right) }\right) \left( 1-\Lambda
^{-\left( 2j+1+s\right) }\right) }, \\
N_{n}^{2} &=&\frac{\Lambda ^{-n\left( 1+s\right) }\left( \Lambda
^{-1}:\Lambda ^{-1}\right) _{n}^{2}}{\left( \Lambda ^{-\left( s+1\right)
}:\Lambda ^{-1}\right) _{n}^{2}\left( 1-\Lambda ^{-\left( 2n+1+s\right)
}\right) },
\end{eqnarray*}%
with }%
\begin{equation*}
\left( a:q\right) _{n}=\left( 1-a\right) \left( 1-aq\right) ...\left(
1-aq^{n-1}\right)
\end{equation*}%
For details, one may refer to Ref.~\cite{Plenio_Chin}.

Then as introduced in ~\cite{VMPS1,VMPS2}, the ground state wave function of
Hamiltonian (\ref{Hamitrans}) can be depicted as
\begin{equation}
\left\vert \psi \right\rangle =\sum_{\{N_{n}\}=1}^{d_{n}}M\left[ N_{1}\right]
...M\left[ N_{L}\right] \left\vert N_{1},...,N_{L}\right\rangle ,
\label{MPSWF}
\end{equation}%
where $N_{n}$ is the physical dimension of each site $n$ with truncation $%
d_{n}$, we employ the standard matrix product representation with optimized
boson basis $\left\vert \widetilde{n}_{k}\right\rangle$ through an
additional isometric map with truncation number $d_{opt}\ll d_{n}$ like in
Refs.~\cite{guo2012critical,Friend} to study the quantum criticality of
ASBM. Each site in the 1D chain can be described by the matrix $M$, which is
optimized through sweeping the 1D chain iteratively to obtain the ground
state, and $D_{n}$ is the bond dimension for matrix $M$ with the open
boundary condition, bounding the maximal entanglement in each subspace.

For the data presented below, we typically choose the same model parameters
in Ref. \cite{guo2012critical,wangyz}, as $\Delta =0.1$, $\omega _{c}=1$, $%
\epsilon =0$, the logarithmic discretization parameter $\Lambda =2$, the
length of the semi-infinite chain $L=50$, and optimized truncation numbers $%
d_{opt}=12$. In addition, we adjust the bond dimension as $D_{max}=20,40$
for $s=0.3,0.7$, respectively, which is sufficient to obtain the converged
results.

\textsl{MCS ansatz.-}. We also apply the MCS ansatz \cite{Ren,mD1,multi_CS2}
to the ASBM. To facilitate the variational study and visualize the symmetry
breaking explicitly, we rotate  the Hamiltonian (\ref{Hamiltonian2}) with $%
\epsilon =0$ around the y axis by an angle $\pi /2$, and have
\begin{eqnarray}
H^{T} &=&-\frac{\Delta }{2}\sigma _{x}+\sum_{k}\omega _{k}a_{k}^{\dag }a_{k}+%
\frac{1}{2}\sum_{k}g_{k}\left( a_{k}^{\dag }+a_{k}\right) \sigma _{z}  \notag
\\
&&+\frac{\lambda }{2}\sum_{k}g_{k}\left( a_{k}-a_{k}^{\dag }\right) i\sigma
_{y}.  \label{Hamiltonian3}
\end{eqnarray}%
The trial state $|\psi ^{T}\rangle $ is written in the basis of the spin-up
state $|\uparrow \rangle $ and spin-down state $|\downarrow \rangle $
\begin{equation}
|\psi ^{T}\rangle =\left(
\begin{array}{c}
\sum_{n=1}^{N_{c}}A_{n}\exp \left[ \sum_{k=1}^{L}f_{n,k}\left( a_{k}^{\dag
}-a_{k}\right) \right] |0\rangle \\
\sum_{n=1}^{N_{c}}B_{n}\exp \left[ \sum_{k=1}^{L}h_{n,k}\left( a_{k}^{\dag
}-a_{k}\right) \right] |0\rangle%
\end{array}%
\right) ,  \label{VM_wave}
\end{equation}%
where $A_{n}$\ ($B_{n}$) is related to the occupation probability of the
spin-up (spin-down) state in the $n$th coherent state; $N_{c}$ and $L$ are
numbers of coherent states and total bosonic modes, respectively; and $%
f_{n,k}$ ($h_{n,k}$) represents bosonic displacement of the $n$th coherent
state and $k$ th bosonic mode. The symmetric MCS ansatz ($A_{n}=\pm {B_{n}}$
with $\pm $\ denotes the even and \ odd parity and $f_{n,k}=-h_{n,k}$) can
only be applied to the delocalized phase, so one can easily detect the
symmetry breaking.

The energy expectation value can be calculated as follows
\begin{equation}
E=\frac{\langle \psi ^{T}|H^{T}|\psi ^{T}\rangle }{\langle \psi ^{T}|\psi
^{T}\rangle }.  \label{ehd}
\end{equation}%
Minimizing the energy expectation value with respect to variational
parameters yields the self-consistent equations, which in turn give the
ground-state energy and the wave function. It has \ been demonstrated that
this wave function with at least a hundred of coherent states can describe
the localized phase of the SBM \cite{Blunden}.

For both VMPS and MCS approaches described above, discretization of the
energy spectrum of the continuum bath should be performed at the very
beginning in the practical calculations. The same logarithmic discretization
is taken for both approaches if comparisons are made.

Within the ground-state wavefunction, the average magnetization $\left\vert
\left\langle \sigma _{x}\right\rangle \right\vert $ is easily calculated.
Note that it can be regarded as the order parameter in the ASBM. The
information of the ground-state can  also be described by the von Neumann
entropy {\ $S_{E}$ of the ASBM, which characterizes the entanglement between
spin and the bosonic bath }%
\begin{equation}
{{S_{E}}=-Tr\left( {{\rho _{spin}\log \rho _{spin}}}\right) },
\label{entropy}
\end{equation}%
{where $\rho _{spin}$ is the reduced density matrix for the spin.} We will
calculate these two quantities together with the average parity to study the
criticality of the ASBM in this work.

\section{Results and discussions}

\subsection{The phase diagram}

\begin{figure}[tph]
\centerline{\includegraphics[scale=0.4]{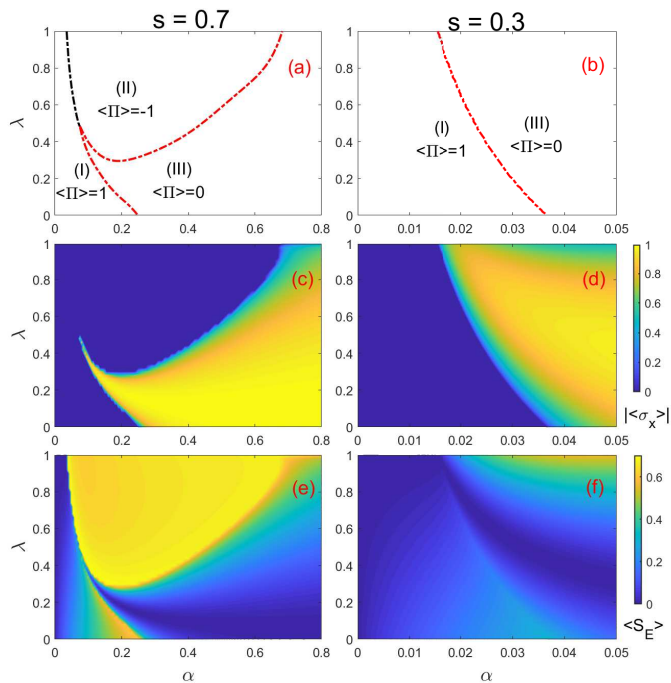}}
\caption{ (Color online) (upper panel) Phase diagram in the $\protect\alpha -%
\protect\lambda $ plane for the ASBM drawn from the Parity $\left\langle \Pi
\right\rangle $: delocalized phases with even (I) and odd parity (II) with
conserved $Z_{2}$ symmetry, and the localized phase (III) with broken $Z_{2}$
symmetry. (middle panel) Order parameter $\left\vert \left\langle \protect%
\sigma _{x}\right\rangle \right\vert $. (lower panel) Entanglement entropy $%
\left\langle S_{E}\right\rangle $. The power of the spectral function is
(left) $s=0.7$ and (right) $0.3$. $\Delta =0.1$, $\protect\omega _{c}=1$.
The parameters used in the VMPS approach are $\Lambda =2$, $L=50$, $%
d_{opt}=12 $, and $D=20,40$ for $s=0.3,0.7$, respectively. }
\label{phase_diagram}
\end{figure}
Generally, the isotropic subohmic SBM exhibits the mean-field critical
behavior for small $s$, and the nonclassical one for large $s$, so we focus
on two typical powers of the spectral function $s=0.7$ and $0.3$ in this
paper. The main results for the ASBM based on  VMPS approaches are
presented in Fig. \ref{phase_diagram} at $s=0.7$ (left) and $0.3$ (right).
The different phases in the ASBM can be precisely characterized by the
parity, which allows for composing the ground-state phase diagrams in the
anisotropy $\lambda $ and the coupling strength $\alpha $ plane in the upper
panel. We call phases I and II  the two delocalized ones with $\left\langle
\Pi \right\rangle =\pm 1$, respectively, and phase III as the localized
phase with $\left\langle \Pi \right\rangle =0$. The boundary between the
phase I and II is marked with the black dashed line, and the boundary of the
phase III and any delocalized phases is indicated with the red dashed line.
Clearly for $s=0.7$, we do observe three phases full with the phase diagram
and a QuTP is the intersecting point of the three critical lines.

Color plots for the order parameter $\left\vert \left\langle \sigma
_{x}\right\rangle \right\vert $ and the entanglement entropy {$S_{E}$}
between the two-level system and the environment bath are displayed in the
middle and lower panels of Fig. \ref{phase_diagram}, respectively. It is
remarkable to see that the skeleton of the phase diagram can be directly
obtained from the color plot of the entropy.  In the small $\alpha$ region,
the entropy increases quickly but still continuously within the phase I area,
and no phase transition takes place.
The order parameter share the
common shape with the phase boundary marked by the red dashed line.

For small power of the spectral function, as shown in  Fig. \ref{phase_diagram} (b) for $s=0.3$ that the phase diagram only consists
of two phases (I and III). It follows that only a single 2nd-order QPT from
the delocalized to localized phase is observed in this case, similar to the
isotropic SBM. This phase diagram can be replotted in a similar way as Fig.
2 in Ref. ~\cite{ciute} for the anisotropic Dicke model, if using $%
\Omega_{E}\propto\sqrt{\alpha}, \Omega_{M}\propto\lambda\sqrt{\alpha}$.

Surprisingly, for large power $s=0.7$, a new delocalized phase with odd
parity (phase II) can grow at the phase III region and have a common border
with phase I, as exhibited in  Fig. \ref{phase_diagram} (a). It intervenes between phases I and III in an unusually way.
The QPT between the two delocalized phases is of 1st-order due to the level
crossing caused by the different wavefunctions with opposite parities,
whereas the QPTs from any delocalized phase to localized phase are
definitely of the 2nd-order due to the symmetry breaking.

For the highly anisotropic case, {both the 1st- and 2nd-order QPTs take
place successively from phases I to II, then to phase III}, similar to the
SBM in the RWA. Note however that the total excitation in the ASBM is not
conserved, unlike the SBM in the RWA. Especially in the moderate anisotropic
model, {with increasing coupling strength, the system would undergo the
2nd-order QPTs for three times: }${I}\rightarrow III$, ${III}\rightarrow II$%
, and $I{I}\rightarrow III$. This unusual reentrance to the same localized
phase has never been reported previously in the light-matter interacting
systems. For the low anisotropic case, since the rotating-wave terms and the
counterrotating terms are comparable, it is not essentially different from
the isotropic SBM, and thus exhibits the similar critical phenomenon.

To be more complete, we have performed extensive calculation based on VMPS for
many different values of $s$, and found that phase II only emerges for $s>0.38$ in the ASBM.
We also extend the anisotropic constant regime to $\lambda >1$%
. In this case, we can absorb $\lambda $ into $g_{k}$, and have $%
g_{k}^{\prime }=\lambda g_{k}$. The coupling strength in Eq. (\ref{spectra})
becomes $\alpha ^{\prime }=\lambda ^{2}\alpha $. Set $\lambda ^{\prime
}=1/\lambda $.  The transformed Hamiltonian is the same as Hamiltonian (\ref%
{Hamiltonian2}) if interchanging the coefficients of the last two terms and
removing primes in $\lambda ^{\prime }$ and $\alpha ^{\prime }$. Regarding $%
\left\vert \left\langle \sigma _{y}\right\rangle \right\vert $ as the order
parameter, the phase diagram can be also obtained in the $\alpha^{\prime}-\lambda ^{\prime
}$ plane for $0<\lambda ^{\prime }<1$, which is  exactly the
same as Fig. 1 for the same $s$. The results for $\lambda >1$ are
qualitatively the same as those for the anisotropic parameter $1/\lambda$ at the same power $s$, because only the phase boundaries are scaled by $1/\lambda^2$ .

To study the QPTs deeply, we will discuss the order parameter and the
entanglement entropy in detail in the next subsections. For more clarity, we
extract the data of the parity, magnetization, and the entropy as a function
of coupling strength $\alpha $ at $\lambda =0.3$ and $0.9$ in Fig. \ref%
{phase_diagram}, and re-plot them in Fig. \ref{QPT_s0.7} for $s=0.7$ and
Fig. \ref{QPT_s0.3} for $s=0.3$, respectively.

\subsection{Order parameter}

\begin{figure}[tph]
\centerline{\includegraphics[scale=0.4]{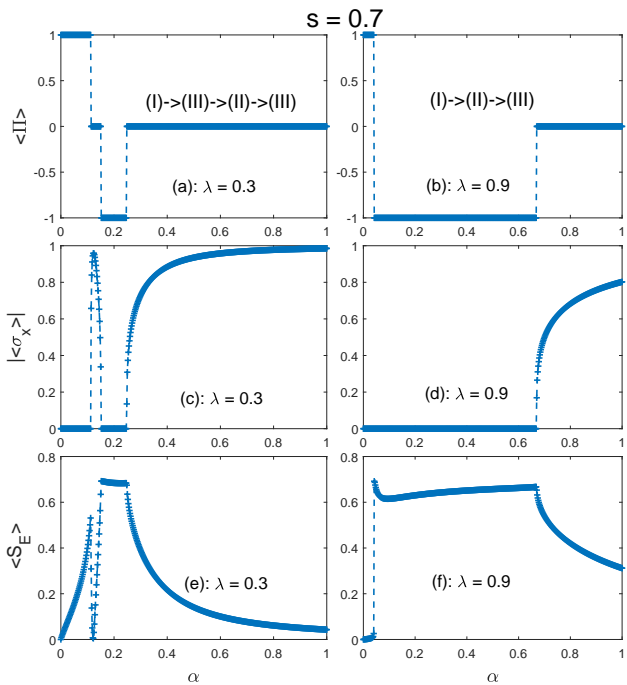}}
\caption{ (Color online) Parity $\left\langle \Pi\right\rangle$ (upper
panels), magnetization $\left\vert \left\langle \protect\sigma %
_{x}\right\rangle \right\vert $ (middle panels), and entanglement entropy $%
\left\langle S_{E}\right\rangle$ (lower panels) as a function of $\protect%
\alpha $ in the ground state for $\protect\lambda = 0.3$ (left) and $\protect%
\lambda = 0.9$ (right) by VMPS approach. $\Delta =0.1$, $\protect\omega %
_{c}=1$, $\protect\epsilon =0$, $\Lambda =2$, $L=50$, $d_{opt}=12$, and $D =
40$ for $s = 0.7$. }
\label{QPT_s0.7}
\end{figure}

\begin{figure}[tph]
\centerline{\includegraphics[scale=0.4]{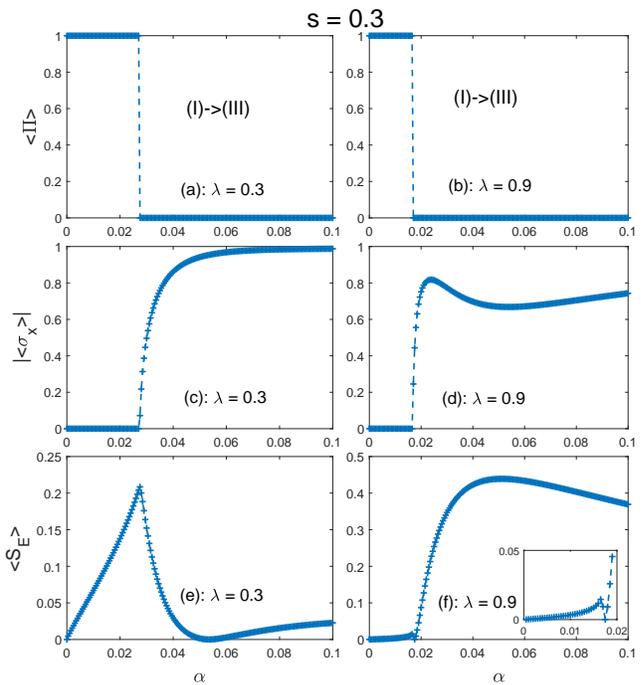}}
\caption{ (Color online) Parity $\left\langle \Pi \right\rangle $ (upper
panels), magnetization $\left\vert \left\langle \protect\sigma %
_{x}\right\rangle \right\vert $ (middle panels), and entanglement entropy $%
\left\langle S_{E}\right\rangle $ (lower panels) as a function of $\protect%
\alpha $ in the ground state for $\protect\lambda =0.3$ (left) and $\protect%
\lambda =0.9$ (right) by VMPS approach. $\Delta =0.1$, $\protect\omega %
_{c}=1 $, $\protect\epsilon =0$, $\Lambda =2$, $L=50$, $d_{opt}=12$, and $%
D=20$ for $s=0.3$. }
\label{QPT_s0.3}
\end{figure}
Generally, in the delocalized phase, spin has  equal probability in the
two states, spin-up and spin-down (both in $x$-axis here), while in
localized phase, spin prefers to stay in one of the two states. Because
phases I and II are delocalized ones with opposite parities ($\pm 1$), the
order parameter must be zero due to the conserved symmetry. So we cannot distinguish phase
II from phase I by the order parameter, which is shown in the blue regime of
Fig. \ref{phase_diagram} (c). Nonzero order parameter
is only found in the localized phase due to symmetry breaking. Thus, $\left\vert \left\langle \protect\sigma_{x}\right\rangle \right\vert $, the order parameter, can only be used to determine the boundary of the continuous QPTs. The  parity
always jumps to different plateaus when crossing any phase boundaries. These
characteristics are clearly shown in the upper panels of  Figs.~\ref{QPT_s0.7}
and {{\ref{QPT_s0.3}, which can be used to compose the phase diagram}}
precisely.

One can indeed see that the order parameter remains zero in  phases I and
II and only becomes nonzero in the phase III in the middle panels of Figs.~%
\ref{QPT_s0.7} and {{\ref{QPT_s0.3}}}. The remarkable peak of the order
parameter in  Fig.~\ref{QPT_s0.7} (c) for $s=0.7,\lambda =0.3$
is originated from the narrow localized phase III.

{\ For small power of the spectral function, {\ say $s=0.3$, there} only
exist two phases: delocalized phase with even parity I and localized phase
III. Although the phase II does not show up in the phase diagram in this
case, it still plays some roles.} The {magnetization for different
anisotropy shows different behaviors after the critical point in the middle
panel of Fig. {\ref{QPT_s0.3}} for $\lambda =0.3 $ and $\ 0.9$. For $\lambda
=0.3$, the order parameter increases monotonously to the global maximum,
while for $\lambda =0.9$, it displays a nonmountainous behavior with $\alpha
$. One can find in the phase diagram that the high anisotropy }$\lambda ${\
and large }$s${\ favor for the emergence of phase II. Even for small $s$, phase
II finally disappears due to the failure in the competition with phase III,
but its effect would not disappear completely without a trace. According to
the different symmetry, it is to note that phase III enhances but phase II
suppresses the magnetization, which cooperate to result in the local minimum
of the magnetization in this region. Of course, if phase II somehow truly
appears in this region, the magnetization must be zero, no local minimum can
be seen.}

\subsection{Entanglement Entropy}

{\ The entanglement entropy $S_{E}$ is presented in the lower panel of Fig. %
\ref{phase_diagram} for }$s=0.7$ and $0.3$. From {the low panels of Figs. 2
and 3, we} can observe that the entropy changes drastically when crossing
all 1st- and 2nd-order critical lines. As shown in Ref. \cite{gu2004} in the
fermionic systems, the entanglement can be used to identify QPTs. \ So, the implications between the entanglement and the quantum
phase in the present ASBM should  also be nontrivial.

To shed some insights, we first consider the 1st-order QPT in the SBM in the
RWA ($\lambda =1$) \cite{wangyz}. In this case, the total excitation $\hat{N}%
=\sum_{k}a_{k}^{\dag }a_{k}+\sigma _{+}\sigma _{-}$ is the conserved number.
At the weak coupling, $\left\langle N\right\rangle =0$, corresponding to
even parity $\left\langle \Pi \right\rangle =1$, the ground state
wave-function is $\left\vert \psi _{0}\right\rangle =\left\vert
0\right\rangle \left\vert \downarrow \right\rangle $ with energy $E_{0}=-%
\frac{\Delta }{2}$, then we can obtain the reduced density matrix for the
spin
\begin{equation*}
\rho _{spin}=\left\vert \downarrow \right\rangle \left\langle \downarrow
\right\vert,
\end{equation*}%
and one can easily obtained entropy $S_{E}=0$ from Eq. (\ref{entropy}).

When exceeding the 1st QPT point, $\left\langle N\right\rangle $ jumps to $1$
corresponding to odd parity $\left\langle \Pi \right\rangle =-1$, the ground
state wave-function for the single excitation is
\begin{equation}
\left\vert \psi _{1}\right\rangle =c\left\vert 0\right\rangle \left\vert
\uparrow \right\rangle +\sum_{k}d_{k}{a_{k}}^{\dag }\left\vert
0\right\rangle \left\vert \downarrow \right\rangle,  \label{one_excited}
\end{equation}%
where $c$ and $d_{k}$ are the coefficients for the bosonic vacuum and single
boson number states. On can easily obtain $c^{2}=\left( 1+\left\langle
\sigma _{z}\right\rangle \right) /2$. The reduced density matrix for the
spin is
\begin{equation}
\rho _{spin}=c^{2}\left\vert \uparrow \right\rangle \left\langle \uparrow
\right\vert +\left( 1-c^{2}\right) \left\vert \downarrow \right\rangle
\left\langle \downarrow \right\vert.
\end{equation}

If $\left\langle \sigma _{z}\right\rangle =0$, we obtain the maximum entropy
$S_{E}^{\max }=\log 2=0.693$ from Eq. (\ref{entropy}). In this case, the
probabilities of spin-up and spin-down are equal, corresponding to the
largest entanglement between spin and bath. In the single excitation state $%
\left\langle \sigma _{z}\right\rangle $ is usually small. e.g. it is found
in Fig. 2(b) of our previous work~\cite{wangyz} that \ $\left\langle \sigma
_{z}\right\rangle $ suddenly switches to a small value around $0.3\pm 0.1$
when crossing the 1st-order QPT point. The entropy in the single excitation
state can be larger than $0.6$.

In the presence of the counter rotating wave terms in the ASBM, the total
excitation $\hat{N}$ is no longer conserved. The state with the even parity
at the weak coupling is not $\left\vert \psi _{0}\right\rangle =\left\vert
0\right\rangle \left\vert \downarrow \right\rangle $ any more, the
components with the even number $\hat{N}$ excitations in the states would be
involved gradually with the increase of the coupling strength, so the
entropy increases within phase I, consistent with the numerical calculations
shown in the lower panels of Figs. 2 and 3.

Because phase II is of odd parity, as long as $\lambda \neq 1$, its state is
different from but close to the state (\ref{one_excited}) with a single
excitation. So the entropy is also high in phase II. We indeed find that the
entropy in all phase II regime is high, indicating it is a highly entangled
phase. As shown in   Fig. 1 (e), a highly entanglement
regime appears in the phase II area. In the 1st-order QPT boundary from
phases I and II, the entropy jumps suddenly to a value close to \ $%
S_{E}^{\max }$ in phase II, as is just shown in
Fig. 2 (f) at $s=0.7,\lambda =0.9$.

In the localized phase of the isotropic SBM, Chin \textsl{et al.} found a
monotonic decrease of entanglement above the transition by means of the
nonadiabatic modes \cite{Chin} analytically, consistent with the numerical
calculations~\cite{lehur}. In the present ASBM, this behavior may be
modified due to the competition between the localized phase III and the
hidden phase II, the latter is lacking in the isotropic SBM but still
possibly present in the ASBM under some condition. In the phase III region
of the lower panel of Fig. 3 for $s=0.3$, at $\lambda =0.3$ and $0.9$, we
note that {the entropy decreases first, reaches a local minimum, and
surprisingly rises again when the coupling strength increases further, in
contrast to the isotropic SBM. As discussed in the last subsection, the
phase III competes with phase II in this area and wins finally. In general,
phase III exhibits a finite value of order parameter but weak entanglement
between spin and bosonic bath, while phase II displays high entanglement but
suppresses the order parameter completely. Although phase II cannot appear
finally, it could still be hidden there and enhance the entanglement. The
observed local minimum is just caused by the cooperated effect of the
competition of phases II and III beyond  the weak coupling. We have
confirmed that, in the strong coupling limit, the entropy in all cases must
vanish (not shown here).}

\subsection{Evidence for 1st-order QPT between the phases with opposite
parities by MCS variational studies}

\begin{figure}[tbp]
\centerline{\includegraphics[scale=0.4]{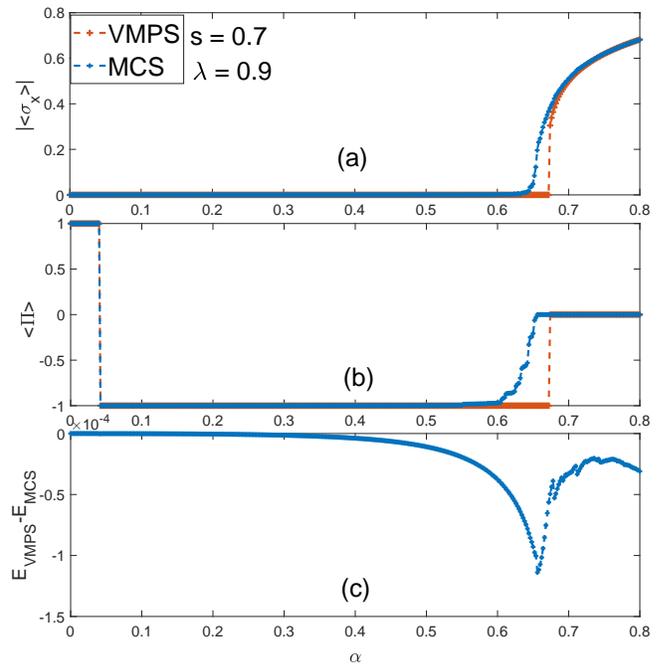}}
\caption{ (Color online) (a) The order parameter $\left\vert \left\langle%
\protect\sigma_{x}\right\rangle\right\vert$, (b) the parity $\left\langle
\Pi \right\rangle$ as a function of the coupling strength within VMPS and
MCS variational approaches. (c) The difference between the VMPS ground-state
energy and that by MCS. $s = 0.7$, $\protect\lambda = 0.9$, $\Delta =0.1$, $%
\protect\omega _{c}=1$, $\protect\epsilon =0$, $\Lambda = 2$, $L = 20$, $%
d_{opt} = 12$, $D = 20$, $N_c = 9$.}
\label{VM}
\end{figure}

The most interesting observation in the ASBM is that a new phase II with odd
parity intervenes between the conventional phase I and III, which is absent
in the isotropic SBM. \ To provide more evidence of this new quantum
phase, we also employ the MCS approach here. {\ By VMPS, for $s=0.7$ and $%
\lambda =0.9$, we have observed that a large region of phase II appears
between  phases I and III. Since all the three phases can be described
well in the trial wave function (\ref{VM_wave}), we in principle can detect
these phases in the MCS framework. In Fig.~\ref{VM}, we list results for the
parity, the magnetization, and the ground-state energy by both MCS and VMPS
approaches for $s=0.7$ and $\lambda =0.9$. Notice that only $L=20$ bosonic
modes are taken for both approaches here due to the computational
difficulties in the MCS approach. However, it does not influence the
essential results at all. The results in the large part of the phase II
regime by both approaches are almost the same, convincingly demonstrating
the existence of the phase II according to its characteristics. The
wavefunction in the MCS reproduces the phase II with the odd parity
explicitly by noting }$A_{n}=-{B_{n}}${. The deviation of the results in the
transition regime between phases II and III is indeed visible, but it does
not influence the existence of phase II. We should point out that the MCS
approach is used here to provide another piece of evidence for the existence
of phase II qualitatively, not for the precise location of the critical
points. }

\subsection{The critical exponent for the order parameters}

\begin{figure}[tbp]
\centerline{\includegraphics[scale=0.4]{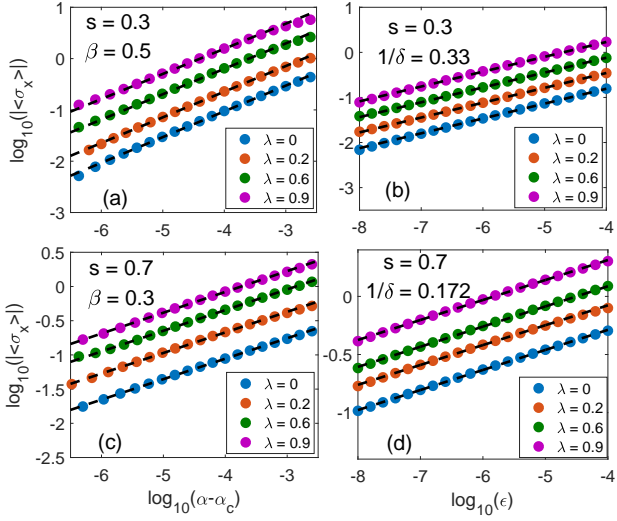}}
\caption{ (Color online) The log-log plot of the magnetization $\left\vert
\left\langle\protect\sigma_{x}\right\rangle\right\vert$ as a function of $%
\protect\alpha-\protect\alpha_c $ (left) at $\protect\epsilon = 0$ and bias $%
\protect\epsilon $ (right) at $\protect\alpha = \protect\alpha_c$ of the
ASBM for $s=0.3$ (upper panel) and $s=0.7$ (lower panel). The numerical
results by VMPS are denoted by blue, orange, green, purple circles for $%
\protect\lambda = 0,0.2,0.6,0.9$ respectively, and the power-law fitting
curves are denoted by the black dashed lines, which shows $\protect\beta =
0.5,0.3$ and $1/\protect\delta = 0.33,0.172$ for $s=0.3,0.7$ respectively.
For visibility, the curves for different $\protect\lambda$ have been shifted
to distinguish them. $\Delta =0.1$,$\protect\omega _{c}=1$, $\Lambda=2 $, $%
L=50$, $d_{opt}=12$, and $D=20,40$ for $s=0.3,0.7$ respectively.}
\label{Fig6}
\end{figure}

The critical behavior of the 2nd-order QPT from phase I to III and from
phases II to III are discussed in this subsection. {\ We present the log-log
plot of the order parameter $\left\vert \left\langle \sigma
_{x}\right\rangle \right\vert $ as a function of $\alpha -\alpha _{c}$ at $%
\epsilon =0$ and as a function of the bias $\epsilon $ at $\alpha =\alpha
_{c}$ for $s=0.3$ and $s=0.7$ with different anisotropic parameters $\lambda
=0,0.2,0.6,0.9$ in the critical regime in Fig. \ref{Fig6}. The order
parameter critical exponents $\beta $ and $\delta $ can be determined by
fitting power-law behavior, $\left\vert {\left\langle {\sigma _{x}}%
\right\rangle }\right\vert \propto {\left( {\alpha -{\alpha _{c}}}\right)
^{\beta }}$ with the bias $\epsilon =0$ and $\left\vert {\left\langle {%
\sigma _{x}}\right\rangle }\right\vert \propto \epsilon ^{1/\delta }$ at $%
\alpha =\alpha _{c} $. All the critical exponents of the anisotropic model
show the same rules as the isotropic SBM. It takes the mean-field value $%
\beta =1/2,1/\delta =1/3$ for $s<1/2$ and the hyperscaling $\beta
<1/2,1/\delta =\left( 1-s\right) /\left( 1+s\right) $ for $s>1/2$. }

\begin{figure}[tbp]
\centerline{\includegraphics[scale=0.4]{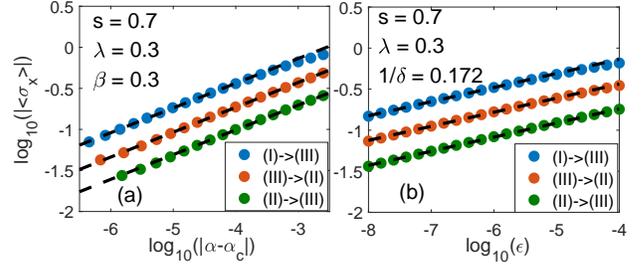}}
\caption{ (Color online) The log-log plot of the magnetization $\left\vert
\left\langle\protect\sigma_{x}\right\rangle\right\vert$ as a function of $%
\left\vert\protect\alpha-\protect\alpha_c\right\vert $ (left) at $\protect%
\epsilon = 0$ and bias $\protect\epsilon $ (right) at $\protect\alpha =
\protect\alpha_c$ of the ASBM for $s=0.7,\protect\lambda = 0.3$. The
numerical results by VMPS are denoted by blue, orange, green, circles for
the critical processes (I) to (III), (III) to (II) and (II) to (III)
respectively, and the power-law fitting curves are denoted by the black
dashed lines, which shows $\protect\beta = 0.3$ (left) and $1/\protect\delta %
= 0.172$ (right) for $0.7$ respectively. For visibility, the curves for
different $\protect\lambda$ have been shifted to distinguish them. $\Delta
=0.1$,$\protect\omega _{c}=1$, $\Lambda=2 $, $L=50$, $d_{opt}=12$, and $D =
40$ for $s = 0.7$ respectively.}
\label{Fig7}
\end{figure}
In the present subohmic ASBM, there is at least one 2nd-order QPT from the
conserved parity phase to the phase III. For large $s$ and the moderate
anisotropy, the model even experiences several 2nd-order QPTs with the
increase of the coupling strength. We also evaluate the critical exponents
for these multiple 2nd-order QPTs for $s=0.7,\lambda =0.3$ in Fig. \ref{Fig7}%
. Very surprisingly, the same critical exponents $\beta $ and $\delta $ are
obtained, indicating that they belong to the same universality class. Based
on these observations, we can say that, counterrotating terms would almost
have no effect on critical exponents even when several 2nd-order QPTs are
present successively at a few critical points for fixed anisotropy in the
ASBM.

The universality in the QuTP in the ASBM is a very challenging issue.
According to the Landau theory, it should be different from those in other
critical points. The numerical calculations cannot be used to isolate the
QuTP from other critical points, and much less distinguish the universality.
The analytical treatment is, however, lacking in any SBMs except in the
ohmic bath, unlike the Dicke models \cite{Dicke,ciute,puhan}. A field theory
formulated from the Feynman path-integral representation of the partition
function for the SBM \cite{Leggett,map,berry,Kirchner} might be extended to
the ASBM. Then analytical arguments based on the quantum-to-classical
mapping would be helpful to clarify this issue.

\section{Conclusion}

We have found rich quantum phases in the ASBM with the subohmic bath by the
VMPS approach. The phase diagram has been composed in the coupling strength
and anisotropy space. \ For large powers of the spectral function, two
2nd-order QPT critical lines meet the 1st-order QPT line at the same point, which is just
a QuTP. At any 2nd-order QPT lines, the critical exponent of the order
parameter and its  field related critical exponents are the same, which
only depend on the power of the spectral function. All phase boundaries can
be precisely determined by the parity and the entanglement entropy, besides,
the 2nd-order QPTs can  also be detected by the magnetization. The 1st-order
QPTs between opposite parity symmetry have been corroborated by the MCS
approach where we can directly observe  opposite parity in the
ground-state wavefunction. For low powers of the spectral function, the
system only experiences the 2nd-order QPT from the delocalized to localized
phases, similar to the isotropic SBM.

The newly found symmetric quantum phase with  odd parity emerges for
large power of the spectra function at the highly anisotropic case and borders to the conventional symmetric phase with even parity, which enriches
the critical phenomena in the spin and boson coupling systems. Although this
phase shares the same odd parity with the phase in the single excitation in
the SBM under the RWA, the total excitation number is not conserved. The QPT
to the localized phase from a delocalized one with odd parity has never been
found before in the spin and boson coupling systems. The ASBM might be
realized in the superconducting circuit QED system where the anisotropic
parameters can be manipulated artificially. We believe that the ASBM would
serve as a new important lab to study the rich quantum criticality.

\textbf{ACKNOWLEDGEMENTS} This work is supported by the National Science
Foundation of China (Nos. 11834005, 11674285), the National Key Research and
Development Program of China (No. 2017YFA0303002),

$^{\ast }$ Email:qhchen@zju.edu.cn


\end{document}